\documentclass[aps,prc,superscriptaddress,reprint,showpacs,showkeys,nofootinbib]{revtex4-1}

\usepackage{epsfig}                   %%
\usepackage{multirow}
\begin{document}
\title{A new study of low-energy (p,$\gamma$) resonances on Magnesium isotopes}

\author{B.~Limata}
    \affiliation{Dipartimento di Scienze Fisiche, Universit\`a di Napoli ''Federico II'', and INFN Sezione di Napoli, Napoli, Italy}
\author{F.~Strieder}
    \email[corresponding author: ]{strieder@ep3.rub.de}
    \affiliation{Institut f\"ur Experimentalphysik, Ruhr-Universit\"at Bochum, Bochum, Germany}
\author{A.~Formicola}
    \affiliation{INFN, Laboratori Nazionali del Gran Sasso (LNGS), Assergi (AQ), Italy}
\author{G.~Imbriani}
    \affiliation{Dipartimento di Scienze Fisiche, Universit\`a di Napoli ''Federico II'', and INFN Sezione di Napoli, Napoli, Italy}
\author{M.~Junker}
    \affiliation{INFN, Laboratori Nazionali del Gran Sasso (LNGS), Assergi (AQ), Italy}
% INFN Sezione di Milano, Italy}
\author{H.W.~Becker}
    \affiliation{Fakult{\"a}t f{\"u}r Physik und Astronomie, Ruhr-Universit\"at Bochum, Bochum, Germany}
\author{D.~Bemmerer}
    \affiliation{Forschungszentrum Dresden-Rossendorf, Bautzner Landstr. 128, 01328 Dresden, Germany}
\author{A.~Best}
	\altaffiliation[present address: ]{Deparment of Physics, University of Notre Dame, Notre Dame, Indiana 46556, USA}
    \affiliation{Institut f\"ur Experimentalphysik, Ruhr-Universit\"at Bochum, Bochum, Germany}
\author{R.~Bonetti}
	\altaffiliation[]{deceased}
    \affiliation{Istituto di Fisica Generale Applicata, Universit\`a degli Studi di Milano and INFN Sezione di Milano, Italy}
\author{C.~Broggini}
     \affiliation{Istituto Nazionale di Fisica Nucleare (INFN), Sezione di Padova, via Marzolo 8, 35131 Padova, Italy}
\author{A.~Caciolli}
     \affiliation{Istituto Nazionale di Fisica Nucleare (INFN), Sezione di Padova, via Marzolo 8, 35131 Padova, Italy}\affiliation{Dipartimento di Fisica, Universit\`a di Padova, Italy}
\author{P.~Corvisiero}
     \affiliation{Universit\`a di Genova and INFN Sezione di Genova, Genova, Italy}
\author{H.~Costantini}
     \affiliation{Universit\`a di Genova and INFN Sezione di Genova, Genova, Italy}
\author{A.~DiLeva}
     \affiliation{Dipartimento di Scienze Fisiche, Universit\`a di Napoli ''Federico II'', and INFN Sezione di Napoli, Napoli, Italy}
\author{Z.~Elekes}
     \affiliation{Institute of Nuclear Research (ATOMKI), Debrecen, Hungary}
\author{Zs.~F\"ul\"op}
     \affiliation{Institute of Nuclear Research (ATOMKI), Debrecen, Hungary}
\author{G.~Gervino}
     \affiliation{Dipartimento di Fisica Sperimentale, Universit\`a di Torino and INFN Sezione di Torino, Torino, Italy}
\author{A.~Guglielmetti}
     \affiliation{Universit\`a degli Studi di Milano and INFN, Sezione di Milano, Italy}
\author{C.~Gustavino}
     \affiliation{INFN, Laboratori Nazionali del Gran Sasso (LNGS), Assergi (AQ), Italy}
\author{Gy.~Gy\"urky}
     \affiliation{Institute of Nuclear Research (ATOMKI), Debrecen, Hungary}
\author{A.~Lemut}
     \altaffiliation[present address: ]{Lawrence Berkeley National Laboratory, Berkley, CA 94720 USA}
     \affiliation{Universit\`a di Genova and INFN Sezione di Genova, Genova, Italy}
\author{M.~Marta}
     \affiliation{Forschungszentrum Dresden-Rossendorf, Bautzner Landstr. 128, 01328 Dresden, Germany}
\author{C.~Mazzocchi}
     \affiliation{Universit\`a degli Studi di Milano and INFN, Sezione di Milano, Italy}
\author{R.~Menegazzo}
     \affiliation{Istituto Nazionale di Fisica Nucleare (INFN), Sezione di Padova, via Marzolo 8, 35131 Padova, Italy}
\author{P.~Prati}
     \affiliation{Universit\`a di Genova and INFN Sezione di Genova, Genova, Italy}
\author{V.~Roca}
     \affiliation{Dipartimento di Scienze Fisiche, Universit\`a di Napoli ''Federico II'', and INFN Sezione di Napoli, Napoli, Italy}
\author{C.~Rolfs}
     \affiliation{Institut f\"ur Experimentalphysik, Ruhr-Universit\"at Bochum, Bochum, Germany}
\author{C.~Rossi Alvarez}
     \affiliation{Istituto Nazionale di Fisica Nucleare (INFN), Sezione di Padova, via Marzolo 8, 35131 Padova, Italy}
\author{C.~Salvo}
     \affiliation{INFN, Laboratori Nazionali del Gran Sasso (LNGS), Assergi (AQ), Italy}
\author{E.~Somorjai}
     \affiliation{Institute of Nuclear Research (ATOMKI), Debrecen, Hungary}
\author{O.~Straniero}
     \affiliation{Osservatorio Astronomico di Collurania, Teramo, Italy}
\author{F.~Terrasi}
     \affiliation{Seconda Universit\`a di Napoli, Caserta, and INFN Sezione di Napoli, Napoli, Italy}
\author{H.-P.~Trautvetter}
    \affiliation{Institut f\"ur Experimentalphysik, Ruhr-Universit\"at Bochum, Bochum, Germany}

\collaboration{The LUNA Collaboration}\noaffiliation
\date{\today}

\begin{abstract}
Proton captures on Mg isotopes play an important role in the Mg-Al cycle active in stellar H shell
burning. In particular, the strengths of low-energy resonances with ${\rm E} < 200$ keV in $^{25}$Mg(p,$\gamma$)$^{26}$Al determine the production of $^{26}$Al and a precise knowledge of these nuclear data is highly desirable. Absolute measurements at such low-energies are often very difficult and hampered by $\gamma$-ray background as well as changing target stoichiometry during the measurements. The latter problem can be partly avoided using higher energy resonances of the same reaction as a normalization reference. Hence the parameters of suitable resonances have to be studied with adequate precision.

In the present work we report on new measurements of the resonance strengths $\omega\gamma$ of the ${\rm E}= 214$, 304, and 326 keV resonances in the reactions $^{24}$Mg(p,$\gamma$)$^{25}$Al, $^{25}$Mg(p,$\gamma$)$^{26}$Al, and $^{26}$Mg(p,$\gamma$)$^{27}$Al, respectively. These studies were performed at the LUNA facility in the Gran Sasso underground laboratory using multiple experimental techniques and provided results with a higher accuracy than previously achieved.

\end{abstract}

\pacs{25.40.Ep, 25.40.Lw, 26.20.Cd, 26.30.-k}
\maketitle

\section{Introduction}                %% Introduction

Observations from satellites \cite{knodlesser99,winkler03} have
mapped the sky in the light of the prominent
$\gamma$-ray line at $E_{\gamma}=1809$ keV of the
$\beta$-decay of $^{26}$Al (T$_{1/2} = 7\times10^5$ yr). The
intensity of the line corresponds to about 3 solar masses of
$^{26}$Al in our galaxy \cite{diehl06}.
Moreover, evidences for an $^{26}$Al excess in the early solar system was found in CAIs (Calcium Aluminum inclusions) showing a significant correlation of $^{26}$Mg (extinct $^{26}$Al) and $^{27}$Al \cite{Lee76,Lee77}.
While the observations from COMPTEL
and INTEGRAL provided evidence that $^{26}$Al nucleosynthesis is
still active on a large scale, the Mg isotopic variations demonstrate that
$^{26}$Mg also was produced at the time of the condensation of the solar-system about 4.6 billion years
ago. Any astrophysical scenario for $^{26}$Al nucleosynthesis must be
concordant with both observations.

The $^{26}$Al is produced mainly via the
$^{25}$Mg(p,$\gamma$)$^{26}$Al capture reaction. The most important
site for the activation of this reaction is the hydrogen-burning
shell (HBS), which may be active in off-main-sequence stars of any
mass \cite{Busso03, Limongi06, Cristallo09}. In particular, the Mg-Al cycle is at work in the hottest
region of the HBS, close to the point of the maximum nuclear energy
release. In the HBS, the $^{25}$Mg(p,$\gamma$)$^{26}$Al reaction starts when the
temperature %\footnote{temperatures are given in ${\rm T}_6 = {\rm T}/10^6$ K}
exceeds about ${\rm T}=30\times10^6$~K and for ${\rm T}=(40 - 60)\times10^6$~K
- corresponding to a Gamow energy of about E$_0 \approx 100$ keV \cite{Rolfs88} -
almost all the $^{25}$Mg is converted into $^{26}$Al.
At higher temperatures, the destruction of $^{26}$Al by
$^{26}$Al(p,$\gamma$)$^{27}$Si and the refurbishment of $^{25}$Mg by
the sequence $^{24}$Mg(p,$\gamma$)$^{25}$Al($\beta^+$)$^{25}$Mg
begins to play a relevant role. The $^{25}$Mg(p,$\gamma$)$^{26}$Al also operates in the carbon and neon burning
shells of massive stars during late stellar evolution.

Moreover, a global anticorrelation between the abundances of Mg and Al has been observed, e.g. in Globular Cluster stars (see \cite{Carretta09} for a recent analysis and references therein). This observation is to the present knowledge coupled to the nucleosynthesis processes involving the Mg-Al cycle occurring in the HBS of primeval generation AGB or massive stars. A detailed knowledge of these processes is a fundamental step toward a
general understanding of the formation of the building blocks of our Galaxy.

The uncertainties in the present stellar models are closely related to a precise evaluation of the relevant reaction rates of the Mg-Al cycle. In particular the reactions $^{24}$Mg(p,$\gamma$)$^{25}$Al and $^{25}$Mg(p,$\gamma$)$^{26}$Al play a key role in those scenarios.

The reaction $^{25}$Mg(p,$\gamma$)$^{26}$Al (Q = 6.306~MeV) is
dominated by narrow resonances. These resonances decay in complex $\gamma$-ray cascades either to the ground state of $^{26}$Al or an isomeric state at ${\rm E_x}=228$ keV. Only the ground state transition is of astrophysical relevance since the ground state decays into the first excited state of $^{26}$Mg with the subsequent $\gamma$-ray emission observed by the satellite telescopes. The isomeric state of $^{26}$Al decays (T$_{1/2} = 6.3$ s) exclusively to the
ground state of $^{26}$Mg and, thus, is not associated with the emission
of $\gamma$-rays. The strengths of these $^{25}$Mg(p,$\gamma$)$^{26}$Al resonances have been experimentally
studied down to an energy\footnote{all energies are given in the center-of-mass frame if not indicated differently}
of E = 190 keV \cite{CH83a,CH83b,EN86,CH86,EN87,EN88,CH89,IL90,EN90,IL96, PO98}.
Nevertheless, the present uncertainty is insufficient for precise models.
In particular, a disagreement between resonance strengths measured by
$\gamma$-ray spectroscopy and delayed AMS (Accelerator Mass Spectrometry) detection of the $^{26}$Al nuclei
after a proton irradiation of $^{25}$Mg at the relevant energies has been reported recently \cite{arazi}.

The nuclear reaction rate of $^{24}$Mg(p,$\gamma$)$^{25}$Al (Q = 2.272~MeV) at astrophysical energies has a contribution by a low-energy resonance at ${\rm E}=214$ keV. Moreover, a strong direct capture component dominates the resonance contribution. The estimate of the latter contribution \cite{HPT74, PO99} is solely based on the experimental data from Trautvetter and Rolfs \cite{HPT74}. Additionally, the ${\rm E}=214$ keV resonance strength carries a large systematic discrepancy between the existing data (e.g. \cite{HPT74, PO99}).

In the present work we report on a new measurement of the strengths of the ${\rm E} = 304$ keV resonance in $^{25}$Mg(p,$\gamma$)$^{26}$Al, as well as the ${\rm E} = 214$ keV resonance in $^{24}$Mg(p,$\gamma$)$^{25}$Al. The radiative capture reaction on the third stable Mg isotope, i.e. the ${\rm E} = 326$~keV resonance in $^{26}$Mg(p,$\gamma$)$^{27}$Al (${\rm Q}=8.272$ MeV), was studied for completeness. These resonances will serve as a normalization for a subsequent determination of astrophysically important low-energy resonances in $^{25}$Mg(p,$\gamma$)$^{26}$Al, i.e. resonance below E$<200$~keV.

The precision and reliability of such normalization standards are important since weak low-energy resonance strengths are often impossible to determine directly from absolute measurements. In particular the target stoichiometry is a critical parameter. Small admixtures of contaminant elements or isotopes in the target, e.g. oxygen as a result of an evaporation process, have already a large effect on the resonance strength determination. Moreover, it is well known in experimental nuclear astrophysics that a solid state target under heavy proton bombardment changes its stoichiometry in the course of the measurement and a frequent control of the target quality is absolutely necessary for long-lasting low-energy measurements. A determination of weak resonance strengths relative to well known resonances can avoid the difficulty of an absolute measurement. The larger yield of high-energy resonances facilitates the determination of the experimental parameters of such resonances. However, these parameters, e.g. target stoichiometry, still need to be measured with high precision: the major goal of the present study.

The resonances were studied using Mg targets with the well known isotopic composition of natural Mg as well as enriched $^{25}$Mg target. The experiments have been performed at the 400~kV LUNA (Laboratory for Underground Nuclear Astrophysics) accelerator in the Laboratori Nazionali del Gran Sasso (LNGS) underground laboratory in Italy \cite{review}. The 1400 m rock overburden (corresponding to 3600 meter water equivalent) of the underground laboratory reduces the $\gamma$-ray background by more than three orders of magnitude for energies higher than 3.5 MeV, compared with a measurement on earth's surface \cite{Bemmerer05}. In order to reduce the systematic uncertainties arising from the detection technique several independent methods have been used. The absolute value of the resonance strengths were measured with both a high resolution HPGe detector and a high efficiency 4$\pi$ BGO summing crystal. The combination of both methods allows for a precise determination of these parameters and the related resonant branching ratios.

As an alternative method - only in case of the  $^{25}$Mg(p,$\gamma$)$^{26}$Al resonance at ${\rm E} = 304$ keV - an enriched Mg target was irradiated with a proton beam and after a proper chemical treatment the number of produced $^{26}$Al nuclei were counted by means of the AMS technique.

In the following sections we will describe in detail the experimental equipment, target preparation and characterization (section \ref{experiment}). The data analysis of the $\gamma$-ray measurements follows in section \ref{analysis} including a description of a {\scriptsize GEANT4} \cite{geant} Monte Carlo code which was used to obtain the efficiency for the 4$\pi$ BGO detector (section \ref{MonteCarlo}). The results of these measurements are given in section \ref{gamma-discussion} and new values for the  weighted average are recommended. A comparison of the $\gamma$-ray measurements with a detection of the reaction products by means of AMS is presented in section \ref{sec-AMS}. Finally, the present work concludes with a discussion and summary of the results (section \ref{discussion}).

\section{Experimental setup and target preparation}\label{experiment}

\subsection{The LUNA accelerator}\label{accelerator}

\begin{figure}
\includegraphics[angle=0,width=\columnwidth]{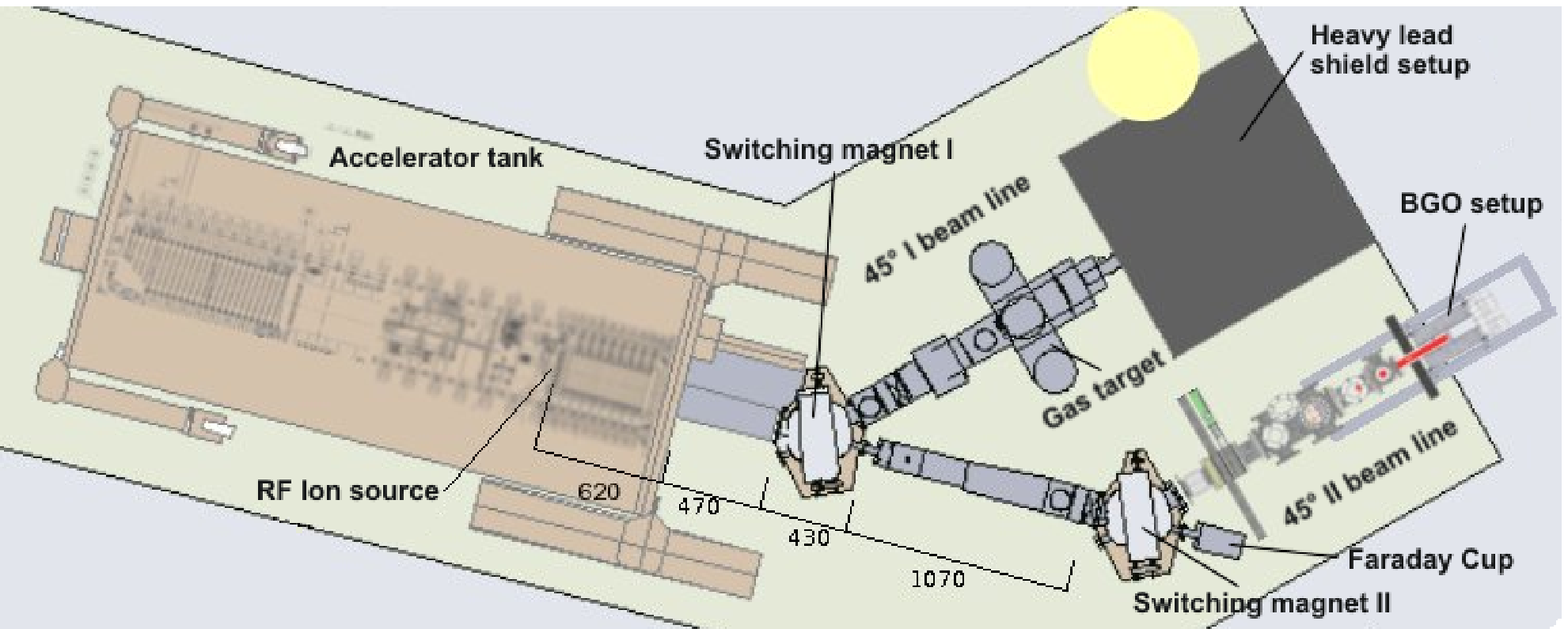}
\includegraphics[angle=0,width=\columnwidth]{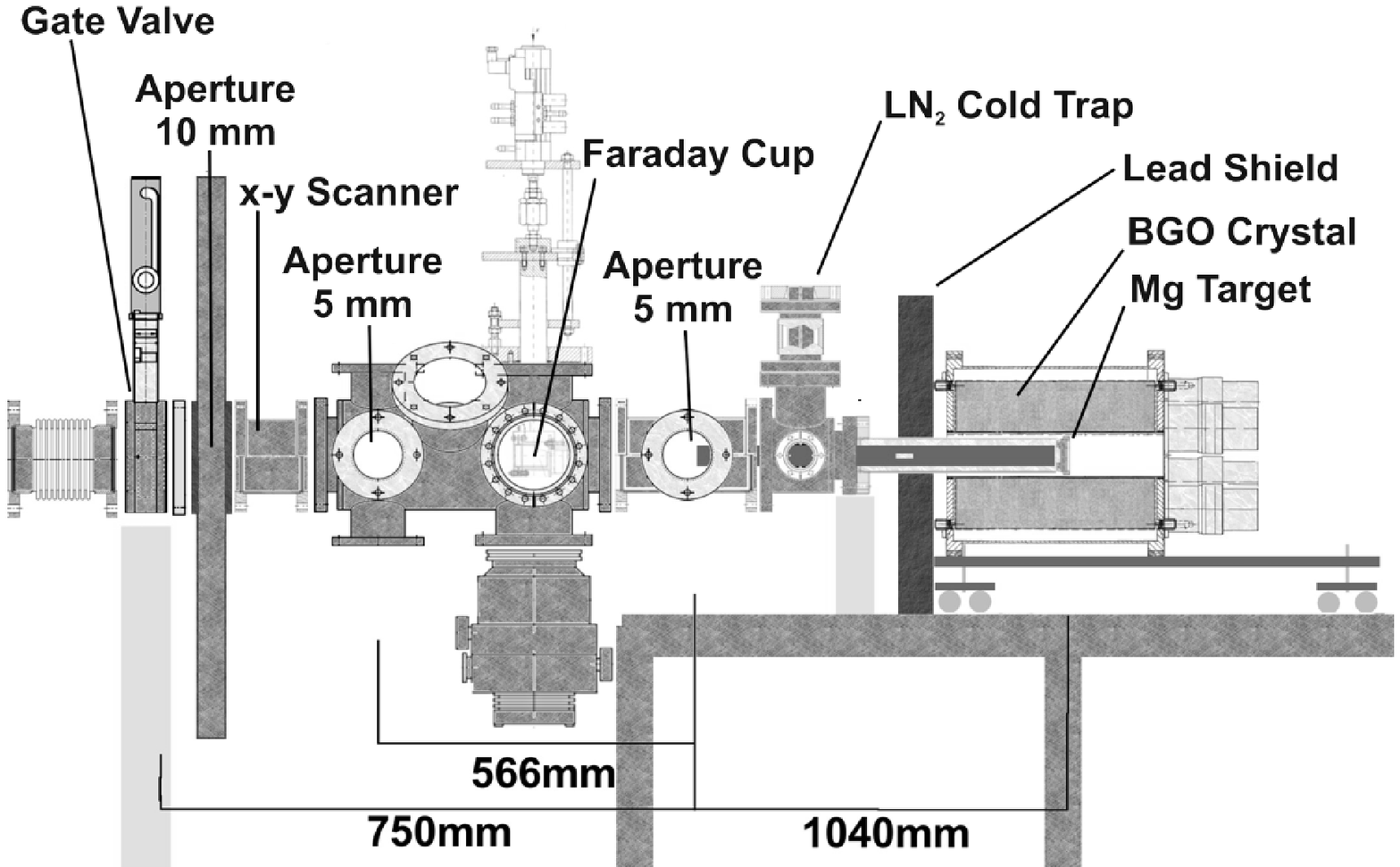}
\caption{(Color online) Floor plan of the 400 kV LUNA accelerator with the 2 beam lines (upper panel) and the $45^\circ$ II beam line with the BGO detector setup (lower panel). All measures are given in mm.}\label{setup}
\end{figure}

The 400~kV LUNA facility has been described elsewhere \cite{Formicola03}. Briefly, the accelerator (Fig. \ref{setup} upper panel) provided in this experiment a proton current on target of up to 250 $\mu$A at energies between E$_p=180$ and 380~keV. The absolute energy is known with an accuracy of 0.3 keV and the energy spread and the long-term energy stability were observed to be 100 eV and 5 eV/h, respectively. The protons are extracted from the radio-frequency ion source and guided under 0$^\circ$ through a vertical steerer and the first 45$^\circ$ switching magnet into a second, identical 45$^\circ$ magnet (distance between both magnets = 1.5 m). With the latter magnet (30 cm radius, 3 cm gap, 1.6 MeV amu) the beam is focussed into the $45^\circ$ II beam line of the LUNA facility. The proton beam passed through a circular, retractable collimator (diameter 10 mm), two focussing apertures (diameter 5 mm each), and a copper shroud ($\ell=1$ m; diameter 28 mm) extending to within 2 mm from the target, where the target plane was oriented perpendicularly to the beam direction. The distance between the two focussing apertures was 566 mm and the second aperture prevented the proton beam from hitting the copper shroud (for details see Fig. \ref{setup} lower panel). The copper shroud was connected to a cold trap cooled to liquid nitrogen temperature. With a turbo pump installed below the cold trap, the arrangement led to a pressure in the target chamber of better than $5\times10^{-7}$ mbar; whereby no C deposition was observed on the targets. A voltage of minus 300 V was applied to the cold trap to minimize emission of secondary electrons from both the target and the last aperture; the precision in the current integration was estimated to be about 2\%. The beam profile on target was controlled by sweeping the beam in the x and y directions within the geometry of the apertures. The targets were directly water cooled in order to prevent any heat damage during the measurements. The BGO detector was mounted on a movable carriage such that the target could be placed in the center of the borehole of the detector maximizing the efficiency of the setup.

\subsection{4$\pi$ BGO summing crystal setup}\label{BGO-exp}

The BGO crystal is a cylinder (length = 28 cm) with a coaxial hole (diameter = 6 cm) and a radial
thickness of 7 cm \cite{Casella02}. The crystal is optically divided in six sectors, each covering a 60$^\circ$ azimuthal angle. In the original configuration two photomultipliers (PMTs) were coupled to the opposite sides of each
sector. In order to allow for a closer distance to the last aperture, all PMTs on one side were replaced with reflecting material. Summing the light produced in all six sectors allows to recover the full energy of detected $\gamma$-rays and, thus, leads to an increased detection efficiency in the case of $\gamma$-ray cascades. Moreover, single spectra can be acquired due to the optical separation of each sector. The energy resolution of each crystal is on the order of 18\% for E$_\gamma=661$ keV.
The signals from the 6 PMTs of the BGO summing crystal were sent to a 16-fold amplifier (CAEN, module N568) which produced, for each incoming pulse, a linear output signal sent to a 12-bit ADC (Silena FAIR, module 9418 V). The
amplifier generated also for each incoming signal a fast output signal. This fast signal generated the acquisition trigger via a constant fraction discriminator (EG\&G, module CF8000) if the fast signal from each PMT is higher than a chosen threshold value. When at least one sector generated a trigger signal, the signals arriving from all the 6
PMTs are converted by the ADC. The total processing time of an event is 24 $\mu$s. The data acquisition is based on a mixed FAIRVME bus \cite{Fair}. The spectra of the BGO sectors were displayed on-line on a PC screen, while the raw data, i.e. the 6 PMT signals for each trigger, were saved event-by-event on a hard disk for an off-line data analysis.

\subsection{HPGe detector setup}\label{HPGe-exp}

In the high resolution phase of the experiment the target holder was replaced by a tube that allowed for an orientation of the target with its normal at 45$^\circ$ with respect to the beam direction. The copper shroud was also cut at 45$^\circ$ such that an evenly distance of 2 mm to the target was ensured. As in the BGO setup the target was directly water cooled. A HPGe detector (115$\%$ relative efficiency, resolution = 2.1~keV at E$_\gamma = 1.3$~MeV) was placed on another moveable carriage oriented at 55$^\circ$ with respect to the beam axis. Thus, target and front face of the detector were not parallel but $\gamma$-ray attenuation effects were reduced compared to the target holder perpendicular to the beam axis and the influence of any angular distributions was minimized. The distance between target and detector could be varied in a range $d=3.5$ to 42.3 cm, where the maximum distance was used for the resonance strength and branching ratio determination. The detector was surrounded by 5 cm of lead, which reduced the background in the low-energy range by a factor 10. Standard electronics was used for processing the detector pulses which were finally stored in a 16k ADC. The acquisition unit was placed close to the experiment and the processed digitized data were sent via Ethernet to a PC for analysis.

\subsection{Target preparation and analysis}\label{target}

A natural Mg target has been produced by evaporation of metallic magnesium of natural isotopic composition on a Ta backing at the IKP of the University of M\"unster, Germany. A small carbon sample was mounted close to the Ta sheet during the evaporation process and later used for an analysis of the target stoichiometry by means of Rutherford Backscattering (RBS). The Mg target on Ta has been cut into two pieces for the HPGe detector and the BGO crystal measurement, respectively.

The RBS analysis of the Mg target on C backing was performed with a 2~MeV He$^+$ beam from the 4~MV Dynamitron-Tandem accelerator of the Ruhr-Universit\"at Bochum, Germany. The beam intensity was about 10 nA and the backscattered particles were detected at an angle of 160$^\circ$ with respect to the beam. The data were analyzed with the computer code RBX \cite{kotai}.

\begin{figure}
\includegraphics[angle=0,width=0.9\columnwidth]{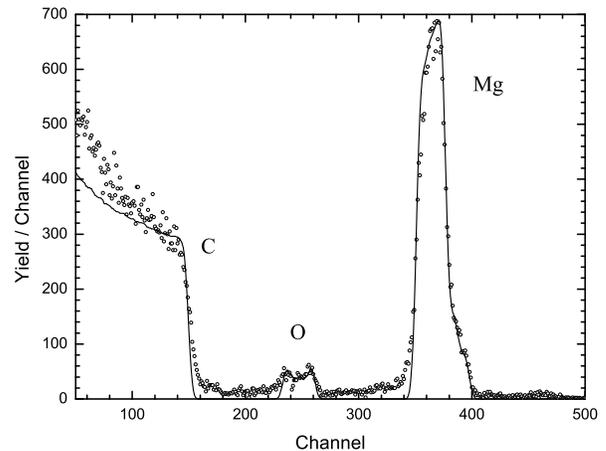}
\caption{Spectrum of the Rutherford backscattering (RBS) analysis of the natural Mg target. The solid line is a fit to the data. The chemical composition of the target and the target layer thickness is derived from this fit.}\label{RBS}
\end{figure}

\begin{figure}
\includegraphics[angle=0,width=0.9\columnwidth]{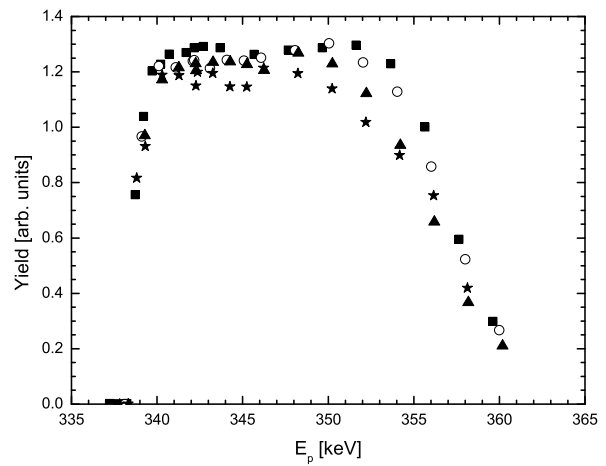}
\caption{Thick-target yield curves obtained for the ${\rm E}=326$ keV resonance of $^{26}$Mg(p,$\gamma$)$^{27}$Al. These data serve as a quality check of the target. The height of the plateau is proportional to the effective stopping power and, therefore, related to the stoichiometry change. The filled squares represent the scan on the fresh target and the open circles, filled triangles, and stars are scans after a charge of 7.2, 8, and 9.6 C, respectively.}\label{target-profile}
\end{figure}

The result of the RBS analysis is shown in Fig. \ref{RBS}. Three regions with different oxygen content can be identified in the target layer. In particular, there is a thin surface layer (thickness = 1.5 $\mu$g/cm$^2$) with a O:Mg ratio of 1:1. In the bulk (thickness = 32 $\mu$g/cm$^2$) of the Mg target layer a ratio of O:Mg = $(0.12\pm0.03)$:1 was found.
Finally, the interface to the backing showed another thin MgO layer which was approximated in the analysis by a thickness of 1.5 $\mu$g/cm$^2$ with O:Mg = 1:1. The structure of this latter layer is probably more complex, but matches sufficiently well the general layering of the target; as a result a slight overestimate of the RBS yield with respect to the data points is observed at the low-energy tail of the Mg peak (Fig. \ref{RBS}). Note, however, that the two thin layers at the surface and the interface have no influence on the resonance strength determination if the resonance energy is locate well inside the Mg bulk. The stoichiometry ratio is independent of stopping power and the uncertainty is mainly based on the quality of the fit. Moreover, the homogeneity of the Mg bulk is demonstrate by the flat thick-target yield plateau of the resonance scans (Fig. \ref{target-profile}).

These stoichiometry results were used in the analysis of the $\gamma$-ray data in order to obtain the effective stopping power (see section \ref{stopping}). The target stoichiometry was checked frequently in close geometry, i.e. thick-target yield curves of each resonance were measured.
Figure \ref{target-profile} shows, as an example, the results of such a scan for the ${\rm E}=326$ keV resonance of
$^{26}$Mg(p,$\gamma$)$^{27}$Al. The maximum observed yield decrease during the course of the experiment was $12\%$. The data were corrected for this stoichiometry change if necessary.

\subsection{Accelerator Mass Spectrometry (AMS)}\label{AMS-intro}

Complementary to the $\gamma$-ray spectroscopy with BGO and HPGe detector the strength of the ${\rm E}=304$ keV resonance of $^{25}$Mg(p,$\gamma$)$^{26}$Al was studied by means of Accelerator Mass Spectrometry (AMS).
The $^{25}$Mg targets for these AMS measurements were prepared at the Laboratori Nazionali di Legnaro, Padova, Italy, by reducing enriched MgO mixed with zirconium powder and evaporation with an electron gun. The target thickness was between 40 and 60 $\mu$g/cm$^2$ corresponding to an energy loss of 20 and 30 keV at the resonance energy. Finally, two targets were analyzed (labelled in the following as A and B, respectively). In parallel to the irradiation of the AMS samples the corresponding $\gamma$-ray yield was observed for each target with the BGO detector in standard geometry. The stoichiometry of the targets could be determined from the $\gamma$-ray yield normalized to the new value of the resonance strength and the O:Mg ratio for the targets A and B turned out to be 0.29 $\pm$ 0.02 and 0.32 $\pm$ 0.02, respectively.

\section{Experimental procedures, data analyses and results of $\gamma$-ray measurements}\label{analysis}

\subsection{Thick-target yield and stopping power}\label{stopping}

The resonance strengths of the resonances at ${\rm E} = 214$, 304, and 326 keV for $^{24}$Mg(p,$\gamma$)$^{25}$Al, $^{25}$Mg(p,$\gamma$)$^{26}$Al, and $^{26}$Mg(p,$\gamma$)$^{27}$Al, respectively, have been measured with both the HPGe and the BGO detector while the $\gamma$-ray branchings can only be determined with the HPGe detector. In particular a precise determination of the resonance strength requires unusual efforts in the measurement of all quantities entering the analysis of this value.
In general the thick-target yield of a narrow resonance is given by the expression \cite{Rolfs88}:
\begin{equation}\label{omegagamma}
Y=\frac{\lambda^2}{2}b_\gamma\omega\gamma\frac{m_{\rm Mg}+m_{\rm p}}{m_{\rm Mg}}\frac{1}{\varepsilon_{\rm eff}}
\end{equation}
where $\omega\gamma$ is the resonance strength, $b_\gamma$ the cross-section fraction that is carried by the
observed $\gamma$-ray (e.g., the branching ratio for a primary transition), $\lambda$ the de Broglie wavelength, and $\varepsilon_{\rm eff}$ the effective stopping power. The latter quantity accounts for the energy loss of the projectiles in the target layer and can be derived with the formula:
\begin{equation}
\varepsilon_{\rm eff}=\varepsilon_{a}+\sum_i\frac{{\rm N}_{i}}{{\rm N}_{a}}\varepsilon_{i}
				\simeq \frac{1}{\rm X_{^yMg}}(\varepsilon_{\rm Mg}+\frac{\rm N_O}{\rm N_{Mg}}\varepsilon_{\rm O})
\end{equation}
with the number of active atoms N$_{a}$ with respect to the inert atoms N$_{i}$, X$_{\rm ^yMg}$ the relative isotopic abundance (or enrichment) of the observed Mg isotope (e.g. ${\rm y}=24$, 25, or 26), and the stopping power $\varepsilon_{\rm O}$ and $\varepsilon_{\rm Mg}$ of protons in oxygen and magnesium at the particular resonance energy, respectively. In this determination of the effective stopping power all other contaminations are neglected since they amount to less than 1\% in total. Therefore, the effective stopping power and, in turn, the $\omega\gamma$ scales with the inverse of the relative isotopic abundance of the effective target isotope and is strongly influenced by the oxygen concentration.

The relative isotopic abundance in case of the natural magnesium target as used for the $\gamma$-ray measurements is well known:
$78.99\pm0.16$\% ($^{24}$Mg), $10.00\pm0.03$\% ($^{25}$Mg) and  $11.01\pm0.03$\% ($^{26}$Mg)
\cite{abundance}. Thus, the effective stopping power for the 3 Mg isotopes including the observed oxygen to magnesium ratio (section \ref{target}) are $\varepsilon_{\rm eff}({\rm ^{24}Mg})=20.7$~eVcm$^2$/10$^{15}$atoms, $\varepsilon_{\rm eff}({\rm ^{25}Mg})=140.2$~eVcm$^2$/10$^{15}$atoms, and $\varepsilon_{\rm eff}({\rm ^{26}Mg})=123.9$~eVcm$^2$/10$^{15}$atoms.
The error of these effective stopping power values is on the order of 4.5~\% each based on the stopping power uncertainties \cite{srim,*SRIMNIMB} of 2.3 and 4.4~\% for protons in oxygen and magnesium, respectively, and including the stoichiometry uncertainty.

\subsection{Measurements with the HPGe detector}

\subsubsection{Efficiency determination}\label{efficiency}

The efficiency of the HPGe detector was studied for different distances $d$ from the target, i.e. 3.5, 8.5, 13.5, and 42.3 cm, with calibrated $\gamma$-ray sources placed at the target position as well as with the ${\rm E} = 259$ keV resonance of $^{14}$N(p,$\gamma$)$^{15}$O \cite{imbriani05}. In addition, the ${\rm E} = 214$ keV resonance of $^{24}$Mg(p,$\gamma$)$^{25}$Al was used for the relative efficiency determination while at the same time in an iterative process the branching ratios for this resonance were improved. In contrast, the reactions $^{25}$Mg(p,$\gamma$)$^{26}$Al and $^{26}$Mg(p,$\gamma$)$^{27}$Al could not be used for such a procedure since the decay schemes of $^{26}$Al and $^{27}$Al are too complex. The absolute scale of the efficiency determination was fixed by the data from the $\gamma$-ray sources, e.g. $^{137}$Cs, $^{207}$Bi, and $^{226}$Ra, while the energy dependence was determined following the approach described in \cite{imbriani05}.
Briefly, the latter procedure is based on the assumption that the intensity ratio of primary and secondary $\gamma$-ray transition for each excited state including the particular detection efficiency must be unity.
These constraints were used in a global fit to the data and the full-energy efficiency $\varepsilon_{\rm FE}({\rm E}_\gamma)$ as a function of $\gamma$-ray energy and distance $d$ was parameterized by the following empirical expression \cite{Gil95}:
\begin{equation}\label{par1}
\varepsilon_{\rm FE}({\rm E_\gamma})=A({\rm E_\gamma},d)\cdot e^{a+b\ln{\rm E_\gamma}+c(\ln{\rm E_\gamma})^2}
\end{equation}
with
\begin{equation}\label{par2}
A({\rm E_\gamma},d) =\frac{1-e^{-\frac{d+d_0}{\alpha+\beta\surd{\rm E_\gamma}}}}{(d+d_0)^2}
\end{equation}
where ${\rm E_\gamma}$ is in MeV, $d$ in cm and $a=8.06\times10^{-2}$, $b=-0.488$, $c=-0.141$, $d_0=0.941$, $\alpha=10.188$ and $\beta=-0.276$, respectively, are fit parameters in the global fit. The summing effect of real coincidences were taken into account similarly as in \cite{imbriani05}.

\begin{figure}
\includegraphics[angle=0,width=0.9\columnwidth]{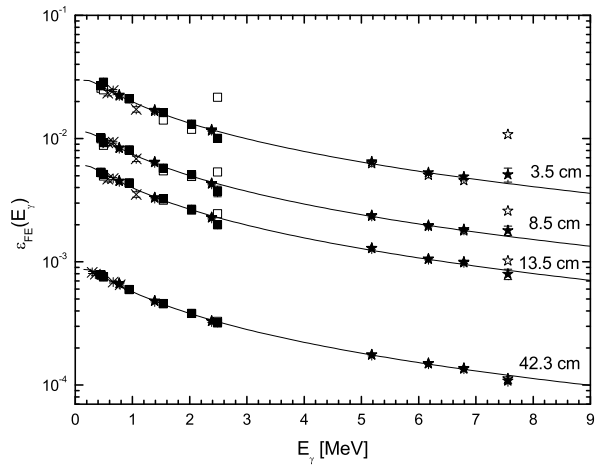}
\caption{Detection efficiency of the HPGe detector as a function of the $\gamma$-ray energy for various distances target - detector front face. The efficiency has been determined with $\gamma$-ray sources (cross) and the reactions $^{14}$N(p,$\gamma$)$^{15}$O (stars) and $^{24}$Mg(p,$\gamma$)$^{25}$Al (squares). The filled symbols denote the data corrected for summing while the open symbols are uncorrected. The solid lines represent the fit curves from eq. \ref{par1}.}\label{effi}
\end{figure}

\begin{table}
\caption{Primary $\gamma$-ray branching ratios of the ${\rm E}=214$ keV $^{24}$Mg(p,$\gamma$)$^{25}$Al resonance from present and previous work.}\label{24Mg}
\begin{ruledtabular}
\begin{tabular}{c c c c}
E$_{X}$ & present work [\%] & \cite{PO99} & \cite{szuecs2010} \\
\hline
$1790$  & $<0.05$       & $<0.8$        & $<0.3$ \\
$1613$  & $<0.05$       & $<0.8$        & $<0.3$ \\
$945 $  & $15.6\pm0.3$  & $15.6\pm 1.1$ & $15.7\pm0.6$  \\
$452 $  & $81.7\pm1.6 $ & $81.7\pm3.4$  & $81.6\pm1.1$  \\
$0   $  & $2.70\pm0.07$ & $2.7\pm0.3$   & $2.69\pm0.08$
\end{tabular}
\end{ruledtabular}
\end{table}

\begin{figure*}
\includegraphics[angle=0,width=0.9\textwidth]{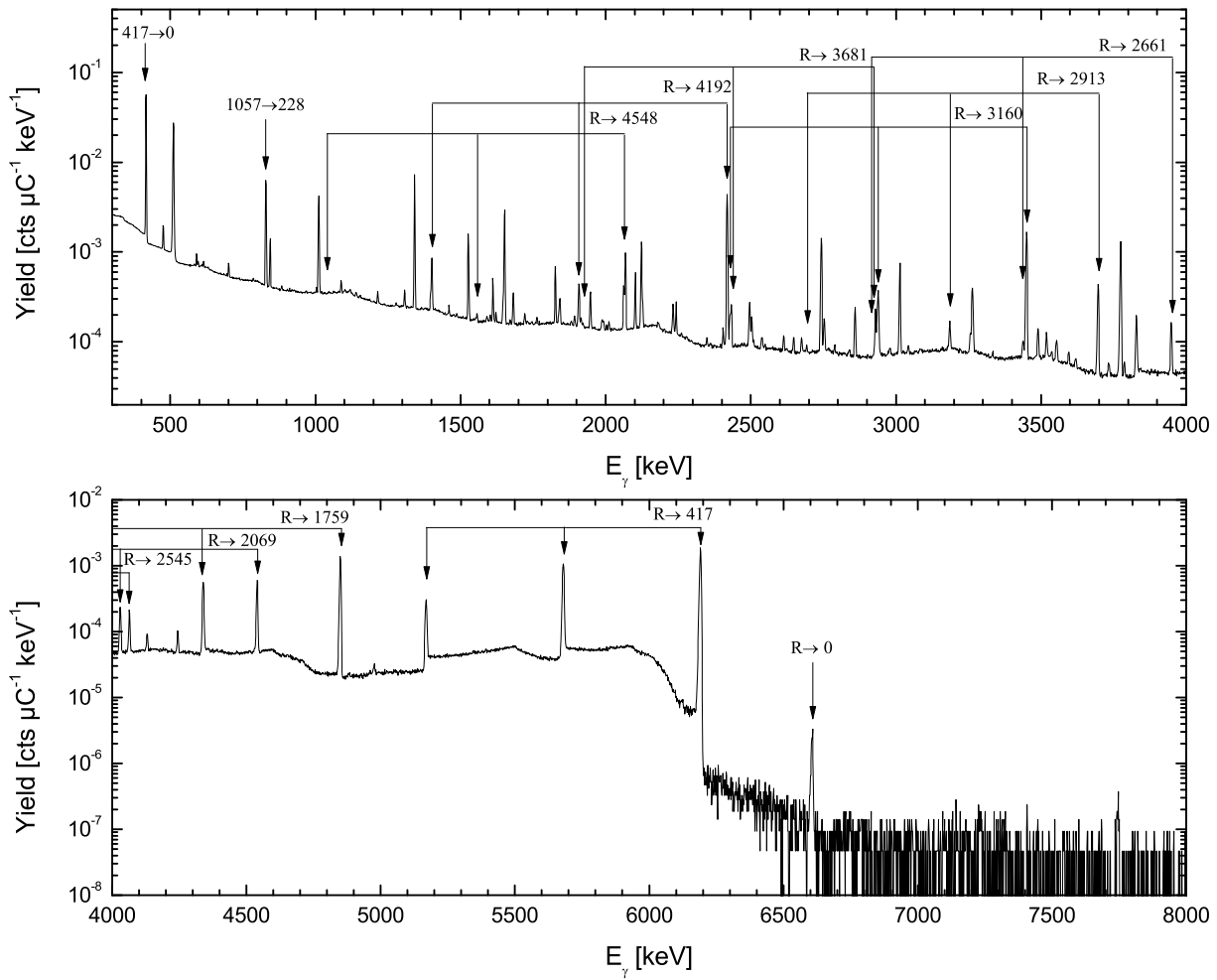}
\caption{The $\gamma$-ray spectrum taken at the ${\rm E}= 304$ keV $^{25}$Mg(p,$\gamma$)$^{26}$Al resonance showing the most prominent primary transitions and some very important secondary transitions.}\label{low-energy-320}
\end{figure*}

Figure \ref{effi} shows the efficiency ${\rm \varepsilon_{FE}(E_\gamma)}$ as a function of the $\gamma$-ray energy for the four distances. In order to illustrate the importance of the summing correction, the open symbols in the figure represent the efficiency values without summing correction while the filled symbols include this correction. Clearly, for distances much larger than 13.5 cm no influence from summing effects is expected. The absolute uncertainty of the efficiency determination is in the order of 3.5\% dominated by the calibration of the radioactive sources. The relative efficiency uncertainty for a measurement at the same distance is lower than 2\%. In particular for the far geometry, $d=42.3$ cm, the efficiency curve is well determined due to the absence of summing effects and the relative uncertainty is below 1.5\%.

\subsubsection{Gamma-ray spectra and branching ratio determination}

The resonances studied in the present experiment give a relatively high yield. As a consequence of the very low $\gamma$-ray background at Gran Sasso these resonances could be observed in far geometry, minimizing the summing effect. For each resonance we could identify all the primary $\gamma$-ray transitions with a very low detection limit, in particular in the case of $^{25}$Mg(p,$\gamma$)$^{26}$Al in the range from 0.02 \% at low-energies, ${\rm E}_\gamma < 1$~MeV, to 0.004 \% for energies above 6.5~MeV. Therefore, the strengths of the resonances and their branching ratios could be determined with high precision. The $\gamma$-ray spectra of %the ${\rm E} = 214$ keV resonance of $^{24}$Mg(p,$\gamma$)$^{25}$Al is shown in Fig. \ref{energy-230},
the ${\rm E} = 304$ keV resonance of $^{25}$Mg(p,$\gamma$)$^{26}$Al is shown in Fig. \ref{low-energy-320} as an example. The spectra for the other resonance are available at \cite{EPAPS}.
%, and the ${\rm E} = 326$ keV of $^{26}$Mg(p,$\gamma$)$^{27}$Al in Fig. \ref{low-energy-340}.
All spectra have been obtained with long runs over several hours - 5 hours minimum and a collected charge of about 7.5 C each - on top of the thick-target yield of each resonance. The background of the $\gamma$-ray lines has been subtracted with off-resonance runs at energies slightly lower than the corresponding resonance energies. For energies above ${\rm E}_\gamma= 4$ MeV the natural background is negligible and the total background is only determined by beam induced background, e.g. from $^{19}$F(p,$\alpha\gamma)^{16}$O.

\begin{table}
\caption{Primary $\gamma$-ray branching ratios of the ${\rm E}=304$ keV $^{25}$Mg(p,$\gamma$)$^{26}$Al resonance.}\label{25Mg}
\begin{ruledtabular}
\begin{tabular}{c c c c }
E$_{X}$ & present work [\%]       & \cite{EN88}\footnotemark[1]    & \cite{IL90}\footnotemark[2]   \\
\hline
$5916$  & $0.09  \pm 0.02 $ & $0.11  \pm 0.02  $ & $               $   \\
$5726$  & $0.10  \pm 0.01 $ & $0.08  \pm 0.02  $ & $0.12\pm 0.03   $   \\
$5457$  &                   & $0.14  \pm 0.02  $ & $               $   \\
$5396$  & $0.22  \pm 0.02 $ & $0.23  \pm 0.03  $ & $0.35\pm0.05    $   \\
$4940$  & $0.08  \pm 0.01 $ & $0.09  \pm 0.02  $ & $               $   \\
$4622$  & $0.28  \pm 0.07 $ & $0.2   \pm 0.04  $ & $0.38\pm  0.06  $   \\
$4599$  & $0.12  \pm 0.01 $ & $0.11  \pm 0.02  $ & $0.13\pm  0.04  $   \\
$4548$  & $1.30  \pm 0.07 $ & $1.13  \pm 0.06  $ & $2.0 \pm  0.1   $   \\
$4349$  & $0.03  \pm 0.01 $ & $0.07  \pm 0.02  $ & $               $   \\
$4206$  & $0.25  \pm 0.02 $ & $0.24  \pm 0.03  $ & $0.25\pm  0.05  $   \\
$4192$  & $19.1  \pm 0.3  $ & $18.7  \pm 0.6   $ & $14.7\pm  0.8   $   \\
$3963$  & $0.17  \pm 0.01 $ & $0.22  \pm 0.03  $ & $0.12\pm  0.05  $   \\
$3750$  & $0.92  \pm 0.02 $ & $0.97  \pm 0.06  $ & $1.5 \pm  0.1   $   \\
$3681$  & $1.09  \pm 0.03 $ & $0.90  \pm 0.08  $ & $0.71\pm  0.08  $   \\
$3675$  & $0.86  \pm 0.13 $ & $0.61  \pm 0.10  $ & $0.59\pm  0.06  $   \\
$3596$  & $4.29  \pm 0.07 $ & $4.2   \pm 0.2   $ & $3.3 \pm  0.2   $   \\
$3160$  & $11.4  \pm 0.2  $ & $11.4  \pm 0.4   $ & $15.6\pm  0.9   $   \\
$3073$  & $0.11  \pm 0.04 $ & $0.18  \pm 0.05  $ & $0.08\pm  0.05  $   \\
$2913$  & $3.04  \pm 0.05 $ & $3.0   \pm 0.1   $ & $4.2 \pm  0.3   $   \\
$2661$  & $1.00  \pm 0.02 $ & $0.97  \pm 0.06  $ & $1.6 \pm  0.1   $   \\
$2545$  & $1.46  \pm 0.03 $ & $1.38  \pm 0.08  $ & $0.9 \pm  0.1   $   \\
$2365$  & $0.47  \pm 0.02 $ & $0.87  \pm 0.19  $ & $0.27\pm  0.07  $   \\
$2069$  & $6.0   \pm 0.1  $ & $5.7   \pm 0.2   $ & $6.5 \pm  0.4   $   \\
$1759$  & $16.1  \pm 0.3  $ & $15.8  \pm 0.5   $ & $22.7\pm  1.3   $   \\
$417 $  & $31.8  \pm 0.5  $ & $33    \pm  1    $ & $24  \pm 1.4    $   \\
$ 0  $  & $0.058 \pm 0.004$ &                    &
\end{tabular}
\end{ruledtabular}
\footnotetext[1]{branchings $<$1\% are given in \cite{Iliadis} as private communication \cite{endt_private}}
\footnotetext[2]{numerical values from \cite{Iliadis}}
\end{table}

The branching ratios for each resonance are given in tables \ref{24Mg}, \ref{25Mg} and \ref{26Mg} for $^{24}$Mg(p,$\gamma$)$^{25}$Al, $^{25}$Mg(p,$\gamma$)$^{26}$Al, and $^{26}$Mg(p,$\gamma$)$^{27}$Al, respectively. Background subtraction and relative efficiency uncertainty were included in the error budget. In all 3 cases the precision of the branching ratios has been improved with respect to the available literature.

\begin{table}
\caption{Primary $\gamma$-ray branching ratios of the ${\rm E}=326$ keV $^{26}$Mg(p,$\gamma$)$^{27}$Al resonance.}
\label{26Mg}
\begin{ruledtabular}
\begin{tabular}{c c c }
E$_{X}$ & present work [\%]       & \cite{IL90}     \\
\hline
$7858$  & $0.09 \pm 0.02 $ & $0.17 \pm 0.03  $    \\
$7280$  & $<0.01         $ & $0.03 \pm 0.01  $    \\
$7071$  & $0.30 \pm 0.02 $ & $0.25 \pm 0.02  $    \\
$6993$  & $0.17 \pm 0.02 $ & $0.20  \pm 0.02  $    \\
$6813$  & $12.1 \pm 0.1  $ & $12.6 \pm 0.7  $    \\
$6776$  & $0.06 \pm 0.01 $ & $0.06 \pm 0.02  $    \\
$6651$  & $0.45 \pm 0.02 $ & $0.50  \pm 0.04  $    \\
$6605$  & $1.26 \pm 0.03 $ & $1.41 \pm 0.09  $    \\
$6158$  & $0.71 \pm 0.03 $ & $0.72 \pm 0.05  $    \\
$6116$  & $0.44 \pm 0.02 $ & $0.34 \pm 0.04  $    \\
$6081$  & $0.59 \pm 0.03 $ & $0.55 \pm 0.05  $    \\
$5752$  & $0.80 \pm 0.03 $ & $0.89 \pm 0.06  $    \\
$5551$  & $2.07 \pm 0.05 $ & $0.39 \pm 0.03  $    \\
$5438$  & $0.22 \pm 0.03 $ & $0.52 \pm 0.04  $    \\
$5248$  & $0.94 \pm 0.03 $ & $0.95 \pm 0.06  $    \\
$5156$  & $0.71 \pm 0.03 $ & $0.03 \pm 0.02   $    \\
$4812$  & $0.54 \pm 0.03 $ & $0.59 \pm 0.05   $    \\
$4410$  & $2.96 \pm 0.07 $ & $3.1  \pm 0.2  $    \\
$4055$  & $10.9 \pm 0.2  $ & $10.7 \pm 0.6   $    \\
$3957$  & $2.64 \pm 0.07 $ & $2.6  \pm 0.2  $    \\
$3680$  & $14.5 \pm 0.2  $ & $13.9 \pm 0.8  $    \\
$2982$  & $19.7 \pm 0.3  $ & $20.2 \pm 0.1  $    \\
$2735$  & $4.43 \pm 0.09 $ & $4.3  \pm 0.3   $    \\
$1014$  & $2.04 \pm 0.07 $ & $2.3  \pm 0.2   $    \\
$844 $  & $19.3 \pm 0.3  $ & $20.2 \pm  0.1    $    \\
$0   $  & $2.06 \pm 0.05 $ & $ 2.5  \pm  0.2     $
\end{tabular}
\end{ruledtabular}
\end{table}

\subsubsection{Resonance strengths}\label{HPGe_strength}

The results are summarized in table \ref{strength}. The uncertainty of the $\omega\gamma$ determination with the HPGe detector is dominated by the absolute error of the $\gamma$-ray efficiency curve (3.5\% for far geometry). A minor contribution arises from the statistical uncertainty which is almost negligible. Common uncertainties of both detection methods, i.e. stopping power and charge integration, are not considered in the error budget of the single measurements, but for the weighted mean of both detection methods (see below). Note in case of $^{25}$Mg(p,$\gamma$)$^{26}$Al we identified transitions feeding either the ground state or the 228~keV isomeric state. The probability for forming the ground state of $^{26}$Al results to $f_0=87.8\pm1.0\%$.

\begin{table*}
\begin{minipage}{\textwidth}
\caption{Resonance strengths of proton captures resonances on magnesium isotopes from the present experiment and previous work.}\label{strength}
\begin{ruledtabular}
\begin{tabular}{c c c c c c c c c c }
  & E [keV]  & \multicolumn{7}{c}{$\omega\gamma$ [meV]} \\
  & & \multicolumn{3}{c}{present work} & \multicolumn{5}{c}{previous work} \\
  & & HPGe\footnotemark[1] & BGO\footnotemark[2] & weighted mean\footnotemark[3] & \cite{angulo} & \cite{IL90}  & \cite{PO99} & \cite{HPT74} & \cite{PO98} \\
\hline
$^{24}$Mg(p,$\gamma$)$^{25}$Al & 214 & $10.4\pm 0.4$ & $11.1\pm 0.6$ & $10.6\pm0.6$ &$10\pm2$ &  &$12.7\pm0.9$ & $10.1\pm2.0$\footnotemark[4] & \\
$^{25}$Mg(p,$\gamma$)$^{26}$Al & 304 & $30.7\pm 1.1$ & $30.6\pm 1.3$ & $30.7\pm1.7$ & $31\pm2$ & $30\pm4$  &  &  & \\
$^{26}$Mg(p,$\gamma$)$^{27}$Al & 326 & $276\pm11$  & $272\pm12$  & $274\pm15$ & $590\pm10$ & $250\pm30$   &  & & $273\pm13$ \\
\end{tabular}
\footnotetext[1]{the uncertainty takes into account the statistical error and a 3.5\% error for the efficiency}
\footnotetext[2]{the uncertainty takes into account the background correction, decay scheme uncertanties and an error for the simulation}
\footnotetext[3]{common uncertainties, i.e. for stopping power and charge integration, are added quadratically}
\footnotetext[4]{original result $\omega\gamma=9.5\pm2.0$ meV corrected for new stopping power data \cite{srim}}
\end{ruledtabular}
\end{minipage}
\end{table*}

\subsection{Measurements with the BGO detector}

\subsubsection{Monte Carlo simulation and efficiency determination}\label{MonteCarlo}

In the BGO setup the Mg target is directly located inside the detector, almost in a $4\pi$ detection geometry. The detector is a high efficiency detection instrument with the disadvantage of the relatively low resolution of the BGO material. Therefore, the BGO detector is the ideal tool to study resonances at lower energies, e.g. ${\rm E}<200$ keV, with small resonance strengths accessible only in a few cases with a HPGe. The advantage of this $4\pi$ geometry is that the influence of any angular distribution and angular correlation effects is strongly reduced compared to smaller detectors.
Moreover, the counting statistics in the $\gamma$-ray spectra were very high even with a reduced proton beam current minimizing the target deterioration, e.g. the change of stoichiometry (see section \ref{target}).
On the contrary, the disadvantage of this approach is that the identification of beam induced background is by far more difficult and in some cases the background lines may be located below the $\gamma$-ray lines of interest.

However, the efficiency determination for the BGO detector is very complex and experimentally almost not accessible. Due to the different multiplicities of each nuclear reaction and the different $\gamma$-ray energies of involved transitions, the total summing efficiency is different for each nuclear reaction. Recently an experimental approach was suggested \cite{spyrou} to first determine these multiplicities, which are then used to derive the corresponding efficiency of the sum peak by means of Monte Carlo simulations. The efficiency determination is simplified in case the multiplicity and the decay scheme are largely known. In the present experiment the efficiency was determined with a Monte Carlo simulation based on {\scriptsize GEANT4} \cite{geant}. The result of the Monte Carlo code is a simulated $\gamma$-ray spectrum. This simulated spectrum can be compared and fitted to the experimental spectrum using only a scaling constant for normalization. The resonance strength can be obtained from the scaling factor and the total event number generated in the Monte Carlo simulation.

The geometry of the BGO detector including beamline, target holder and support structure was implemented in the {\scriptsize GEANT4} code. During the initialization of the code the branching ratios and $\gamma$-ray energies of the selected resonance are loaded. This includes not only the primary transitions - taken from the HPGe phase of the present work - but also all relevant secondary transitions. All available information have been used to construct the full decay schemes of the resonances. Thus, the complete $\gamma$-ray deexcitation of the compound nucleus is followed down to the ground state, in case of $^{26}$Al also to the isomeric state at E$_x=228$ keV.
For each excited state a random number generator selects the subsequent excited state and, hence, the emitted $\gamma$-ray energy according to the implemented feeding probability. In some cases up to six different $\gamma$-rays are emitted per simulated event: the multiplicity of the event.

The point of origin of the $\gamma$-ray emission in the simulation is located on the target and the $\gamma$-rays are tracked through the geometry of the setup. In case the $\gamma$-ray deposits energy in the active volume of the BGO crystal the particular energy loss is stored in a histogram. In this way a single spectrum from each of the 6 segments as well as the sum spectrum of the full event is constructed. The energy resolution of each single spectrum was adapted separately to the experimental energy resolution of the BGO sectors.
Angular distribution or angular correlation effects have not been taken into account in the simulation. In order to allow for a full analysis of the experimental spectra also simulations for background reactions like $^{11}$B(p,$\gamma$)$^{12}$C, $^{18}$O(p,$\gamma$)$^{19}$F or $^{19}$F(p,$\alpha\gamma$)$^{16}$O could be obtained from the code.

The efficiency estimate of the simulation was tested with $\gamma$-ray sources placed at the target position, i.e. $^{137}$Cs and $^{60}$Co. The results of measurement and simulation agreed to better than 2\%. Furthermore, the validity of the Monte Carlo code was verified for a different detector setup, i.e the $12\times12$'' NaI detector at the Dynamitron-Tandem Laboratory of the Ruhr-Universit\"at Bochum \cite{Mehrhoff}. The present code delivered the same efficiency curve as a totally independent code based on {\scriptsize GEANT 3.21}. Finally, the results of both codes agreed very well with measurements of various nuclear reactions testing the characteristics of the NaI detector (for details see \cite{Best}). In summary, the efficiency determination of the BGO detector is reliable to better than 3\%.

\subsubsection{Data analysis}

The same natural Mg target was used for all measurements. The running times were $t_{\rm L}=330$, 8150, and 400 s, at the E = 214 ($^{24}$Mg(p,$\gamma$)$^{25}$Al), 304 ($^{25}$Mg(p,$\gamma$)$^{26}$Al), and 326 keV ($^{26}$Mg(p,$\gamma$)$^{27}$Al) resonances, respectively. The dead time was always kept below 4 \%. In between these runs the thick-target yield curve for each resonance was obtained in order to determine the best energy for the measurement. The total charge collected during the course of the experiment on target was less than 0.5 C with an average proton current of 10 - 40 $\mu$A. The target deterioration was checked with resonance scans before and after the measurement and found to be negligible.

The data were stored in an event-by-event mode (list mode) where the energy information and the corresponding crystal segment were recorded. The single spectra for all six BGO segments were extracted from the list mode data and separately energy calibrated.  Thus, the total summing spectrum could be reconstructed from the data after the energy calibrations have been matched.

\begin{figure}
\includegraphics[angle=0,width=\columnwidth]{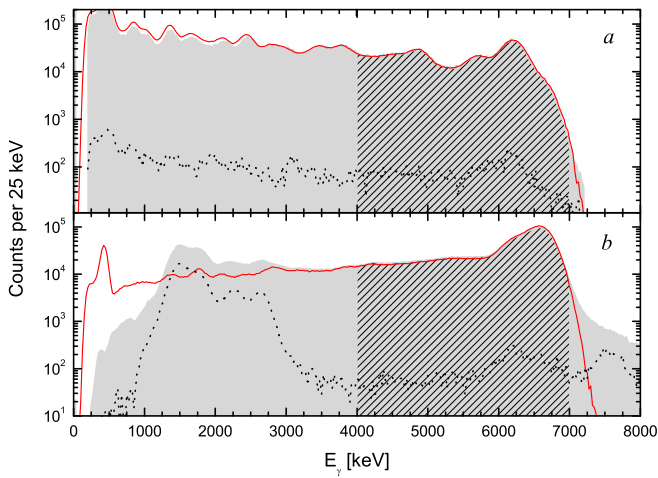}
\caption{(Color online) Gamma-ray spectra taken with the BGO detector at the ${\rm E}= 304$ keV resonance of $^{25}$Mg(p,$\gamma$)$^{26}$Al: $a$) single sum (of all 6 crystals); $b$) total sum spectrum. The shaded area, solid red line, and dotted line represent the measurement, the {\scriptsize GEANT4} simulation, and the background run, respectively. The hatched area illustrates the fitted energy region (for details see text). %All notations as in Fig. \ref{BGO_24Mg}.
}\label{BGO_25Mg}
\end{figure}

Gamma-ray spectra for the case of $^{25}$Mg(p,$\gamma$)$^{26}$Al are displayed in Fig. \ref{BGO_25Mg} (other spectra available at \cite{EPAPS}) both the incoherent sum of the single crystal spectra (a, in the following called single sum) and the total summing spectrum (b, in the following called total sum). A background measurement at a beam energy slightly lower than the corresponding resonance energy and the simulated spectrum is shown for comparison. The $\gamma$-ray energy region used to fit experimental data and simulation is indicated. The reduced yield in the experimental total sum below ${\rm E_\gamma}=1.3$ MeV in on- and off-resonance runs is caused by the energy threshold of the data acquisition trigger on the on-line sum signal (see section \ref{BGO-exp}) and has no effect on the single crystal spectra. This trigger threshold had no impact on the analysis of the $\gamma$-ray spectra. Furthermore, a coincidence condition was applied to all events requiring the full energy being in the indicated energy region or above. As a consequence the environmental background is discriminated and the structure of the decay scheme appeared as simulated. Note that the total sum spectra are more sensitive to pile-up. This effect can be observed on the high-energy tail of all full-energy sum peaks and is probably caused by accidental coincidences with low-voltage (low-energy) noise from the PMTs of the BGO detector. An energy cut in order to discriminate those events could not be applied during the off-line analysis since the real low-energy events are necessary to construct the full-energy event. This effect leads in all cases to a slight disagreement between simulated and experimental spectra on top of the full-energy sum peak. The disagreement is not larger than 3\% of the total number of counts in the region of interest and in agreement with the number of events found in the pile-up region above the full-energy sum peak.

The agreement between the single sum for all 3 reactions and the corresponding simulations is almost perfect and those spectra have been used to obtain the resonance strengths. However, the pile-up effect becomes almost negligible and the analysis of single and total sum agrees to better than 1\% when reducing the proton beam current drastically (below 0.5 $\mu$A). Unfortunately, the current measurement gets unreliable due to the bad focussing properties of the accelerator system at such low intensities and consequently enlarged secondary electron emission: those runs could not be used for absolute measurements.

The $\gamma$-ray background at the ${\rm E}=214$ keV resonance of $^{24}$Mg(p,$\gamma$)$^{25}$Al is dominated by the natural room background. A background measurement acquired directly after the on-resonance run was added to the {\scriptsize GEANT4} simulation after gain and run time matching in order to account for this background component in the analysis. The combined spectrum was then fitted to the on-resonance run.

In contrast to the $^{24}$Mg(p,$\gamma$)$^{25}$Al resonance the measurements of the $^{25}$Mg(p,$\gamma$)$^{26}$Al and $^{26}$Mg(p,$\gamma$)$^{27}$Al resonances were only influenced by beam-induced background. Gamma-ray lines from the reactions $^{19}$F(p,$\alpha\gamma$)$^{16}$O and $^{18}$O(p,$\gamma$)$^{19}$F can be observed in the background spectra. In particular the well known E$_\gamma=6.13$ MeV line of the strong $^{19}$F(p,$\alpha\gamma$)$^{16}$O resonance at ${\rm E}=324$ keV is present in the $^{26}$Mg(p,$\gamma$)$^{27}$Al measurement. Background and on-resonance measurement are both very close to the $^{19}$F(p,$\alpha\gamma$)$^{16}$O resonance energy and, therefore, the yield from this background source is very sensitive to the exact proton energy. As a consequence a background subtraction based on an off-resonance measurement is impossible and the energy region of this $\gamma$-ray line has been excluded from the analysis \cite{EPAPS}.

\begin{figure}
\includegraphics[angle=0,width=\columnwidth]{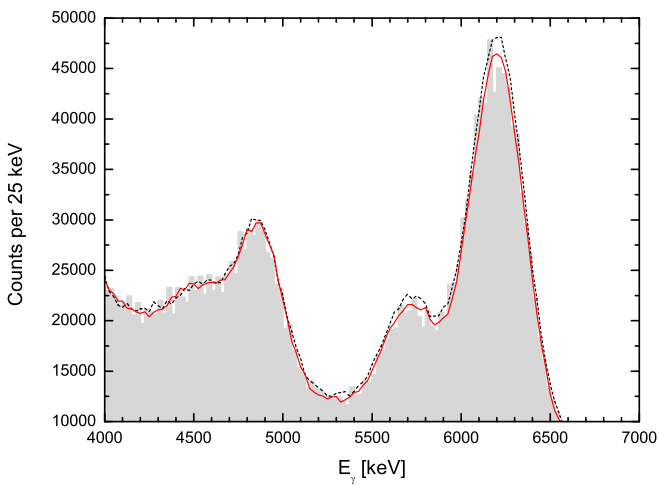}
\caption{(Color online) Single sum $\gamma$-ray spectrum (shaded area) as shown in Fig. \ref{BGO_25Mg} compared to a simulation with the present branching ratio data (red solid line) and from Ref. \cite{EN88} (dashed line).}\label{BGO_comp}
\end{figure}

The branching ratios had to be included in the {\scriptsize GEANT4} code and as a consequence only the resonance strength can be obtained with this detection technique. However, the fit quality is an indication of the branching ratio precision. The effect of different primary branching ratios was tested in case of $^{25}$Mg(p,$\gamma$)$^{26}$Al. In Fig. \ref{BGO_comp} the experimental single sum spectrum is compared to a simulation based on the data of \cite{EN88} and the present work, respectively. The relative intensities of the $\gamma$-ray peaks are reproduced with both data set. Nevertheless, a slightly better fit is achieved with the present data. This agreement demonstrates the internal consistency of all phases of the present experiment and emphasizes the achieved precision. The uncertainty estimated from the influence of the decay scheme is at most 3 \%. The yield in the total sum depends only weakly on these parameters. This is important for weak resonances with largely unknown branching ratios and another advantage of this method.

\subsubsection{Resonance strengths}

The results on the resonance strength are summarized in Table \ref{strength}. The systematic error of $\omega\gamma_{\rm BGO}$ is given by the uncertainty of the simulation, the decay scheme and the background correction. This includes also the uncertainty of the pile-up effect on the total sum.
The measurements of the $^{25}$Mg(p,$\gamma$)$^{26}$Al and $^{26}$Mg(p,$\gamma$)$^{27}$Al resonances are essentially background free and the uncertainty of the simulation may be as large as 3\%. The influence of the background cannot be neglected in case of the $^{24}$Mg(p,$\gamma$)$^{25}$Al resonance and an increased uncertainty of the simulation of 5\% accounts for this issue. The decay scheme uncertainty was discussed in the previous section and estimated to 3\%. The statistical uncertainties of the measurements were in the order of 0.1\% and neglected in the present analysis.

These uncertainties are independent from the HPGe measurement and have to be considered for the calculation of the weighted average of the $\omega\gamma$ values. The additional errors for the stopping power (4.5\%) and the charge measurement (2\%) have to be added quadratically to the error of the weighted mean in order to evaluate a total uncertainty of the present $\gamma$-ray experiment.

Finally, the resonance strengths from the BGO and HPGe measurements are in perfect agreement within their errors. The presented analysis demonstrates that with the BGO detector a similar precision as with a HPGe detector can be reached. The challenging measurements of $^{25}$Mg(p,$\gamma$)$^{26}$Al resonances below ${\rm E} >200$~keV with $\omega\gamma\ll1\mu$eV is feasible with such an approach and will lead to an improved knowledge of the reaction rate at astrophysical energies.

\subsection{Discussion of $\gamma$-ray measurements}\label{gamma-discussion}

\subsubsection{The ${\rm E}=304$ keV resonance in $^{25}$Mg(p,$\gamma$)$^{26}$Al}

The strength of this resonance is the most important parameter in the context of the present work and, therefore, will be discussed first. The weighted average of the present experiment, $\omega\gamma=30.7\pm1.7$ meV, is in very good agreement with previous work, i.e. the latest $\gamma$-ray spectroscopy experiment \cite{IL90}, but also with older experiments \cite{CH83b, EN86}. However, the present result strongly disagrees by more than 3$\sigma$ with the resonance strength, $\omega\gamma=24\pm2$ meV, measured by means of AMS in \cite{arazi} - if scaled for the total $\omega\gamma$. In \cite{arazi} the authors do not prove the stoichiometry of their target and no test for oxygen contamination is presented. From the present work it is evident that such contaminations cannot be neglected and might have an effect of 20\% on the resonance strength. As a consequence the AMS value of \cite{arazi} will not be considered for the further analysis. The NACRE compilation \cite{angulo} provided a weighted average over all published $\gamma$-ray measurements and recommended a value $\omega\gamma=31\pm2$~meV. This recommendation is in perfect agreement with the present result and we combine the NACRE value with the present $\omega\gamma$. This procedure yields a new recommended resonance strength of $\omega\gamma=30.8\pm1.3$~meV for the ${\rm E}=304$ keV resonance of the reaction $^{25}$Mg(p,$\gamma$)$^{26}$Al. Finally, in a recent series of papers \cite{ReacRate1,ReacRate2,ReacRate3,ReacRate4} a new reaction rate compilation is presented. Our present result is also in good agreement with the value, $\omega\gamma=30.0\pm3.5$~meV \cite{ReacRate3}, used in this compilation to calculate the rates, while the uncertainty reflects the previous knowledge of the resonance strength.

\subsubsection{The ${\rm E}=214$ keV resonance in $^{24}$Mg(p,$\gamma$)$^{25}$Al}

The lowest known resonance in $^{24}$Mg(p,$\gamma$)$^{25}$Al at ${\rm E}=214$ keV is less constraint than the reaction $^{25}$Mg(p,$\gamma$)$^{26}$Al and only a few measurements exist. However, the most recent experiment of the TUNL group \cite{PO99} resulted in a 25\% higher resonance strength than recommended by NACRE. The NACRE value is essentially based on the result of Trautvetter and Rolfs \cite{HPT74}. This experiment \cite{HPT74} has to be corrected for updated stopping power data \cite{srim} which leads to a 6\% higher $\omega\gamma$ value, but still only slightly consistent with \cite{PO99} (see Table \ref{strength}). The present result, $\omega\gamma=10.6\pm0.6$~meV, is in good agreement with \cite{HPT74} and differs from \cite{PO99} by about 2$\sigma$. A weighted average of these 3 experiments leads to a recommended value of $\omega\gamma=11.2\pm0.9$ where the uncertainty was determined by the scale factor method \cite{PDG} with an inflation factor of $\sqrt{\chi^2/\chi^2(P=0.5)}$ \cite{SFII} for $\chi^2=4.8$ and 2 d.o.f. %given by an external error due to the disagreement of the single data sets.

\subsubsection{The ${\rm E}=326$ keV resonance in $^{26}$Mg(p,$\gamma$)$^{27}$Al}

The reaction $^{26}$Mg(p,$\gamma$)$^{27}$Al was measured for completeness although its astrophysical relevance is very small. However, the available experimental data - compiled in NACRE \cite{angulo} - differ considerably from each other. Two experiments \cite{Buc80, IL90} result in a value as low as $\omega\gamma=250$ meV while all other experiments \cite{VA56, VA63, KE80, SM82} give a resonance strength around $\omega\gamma\approx700$ meV. Therefore, the recommended value of the NACRE compilation of $\omega\gamma=590\pm10$ meV is questionable and in particular the uncertainty is not justified. The present result of $\omega\gamma=274\pm15$ meV for the ${\rm E}=326$ keV resonance of the reaction $^{26}$Mg(p,$\gamma$)$^{27}$Al is in excellent agreement with the most recent result of \cite{PO98} and in good agreement with the results of \cite{Buc80, IL90}. This comparison suggests in view of the internal consistency of the present experiment that experiments giving a 3-fold higher $\omega\gamma$, e.g. \cite{VA56, VA63, KE80, SM82}, should be discarded. The weighted average of the 3 remaining experiments leads to a recommendation of $\omega\gamma=269\pm10$ meV.

\section{AMS measurement}\label{sec-AMS}

\subsection{Requirements and irradiation}

An AMS measurement provides the ratio between a rare and an abundant isotope in the same sample. Therefore, the reaction $^{25}$Mg(p,$\gamma$)$^{26}$Al represents an ideal case for an AMS study of the resonance strengths. The lowest observable isotopic ratio determines the sensitivity limit and is usually in the order of 10$^{-15}$.  Adding a known amount of stable $^{27}$Al to the sample material after proton irradiation allows for a precise determination of the absolute $^{26}$Al content in the sample from the experimental $^{26}$Al/$^{27}$Al isotopic ratio. Due to the short lifetime of the $^{26}$Al isomeric state, this off-line technique yields directly the astrophysical important ground state contribution to the resonance strength.

The irradiation of the enriched $^{25}$Mg targets (section \ref{AMS-intro}) was carried out in the BGO standard geometry. The capture $\gamma$-ray emission was observed concurrently with the BGO crystal and the ratio between AMS and $\gamma$-ray spectroscopy could be determined directly. Any systematic differences between the two methods as suggested by the measurement of Arazi et al. \cite{arazi} would lead to a ratio different from unity.

The difficulty of such an AMS measurement is a reliable monitoring of the target stoichiometry and quality during the irradiation. In $\gamma$-ray spectroscopy this can be easily achieved by a resonance scan of a well-known resonance at higher energy of the same reaction. In general, in AMS the use of this technique is very limited, often it cannot be applied at all since the procedure would lead to a large amount of additional reaction products, i.e. $^{26}$Al. Fortunately, the observed ${\rm E}=304$~keV resonance in $^{25}$Mg(p,$\gamma$)$^{26}$Al is by itself strong enough that a complete resonance scan could be measure in a relatively short period. The collected charge during the resonance scans was always kept below 0.1\% of the total irradiation charge. Therefore this contribution has been neglected in the analysis. The change of the target stoichiometry during irradiation was taken into account.

In order to study the influence of experimental parameters on the AMS results, e.g. beam power, target A was irradiated with high beam intensity and short irradiation time while target B was irradiated with low intensity over a longer period. A non-irradiated target - target C - served as an AMS blank sample for background reference (Table \ref{AMS-samples}).

\begin{table}
\begin{ruledtabular}
\caption{Experimental parameters of the $^{25}$Mg AMS sample irradiation and the amount of $^{27}$Al added to each sample during the chemical process (for details see text).}\label{AMS-samples}
\begin{tabular}{l c c c c}
AMS sample & target & E$_p$ [keV] & Charge [C] & $^{27}$Al [mg] \\
\hline
304-S1  & A & $321$ & 1.316\footnotemark[1] & 1.0 \\
304-S2  & A & $321$ & 1.316\footnotemark[1] & 1.0 \\
304-S3  & A & $321$ & 1.316\footnotemark[1] & 1.0 \\
304-R1  & B & $322$ & 0.0187 & 0.5 \\
304-BLK & C & $-$ & $-$ & 0.5 \\
\end{tabular}
\end{ruledtabular}
\footnotetext[1]{Total charge collected on target A was 5.264 C, but the extracted material was divided into 4 samples.}
\end{table}
	
\begin{table*}
\begin{minipage}{\textwidth}
\begin{ruledtabular}
\caption{Results of the $^{26}$Al/$^{27}$Al ratios determination with AMS. The values have been obtained from two different measurement periods with independent reference checks as listed.}\label{AMS_results}
\begin{tabular}{l c c c c c c c}
sample & t$_{\rm AMS}$ [s] & ${\rm \bar I}_{\rm ^{27}Al}$ [pnA]  & N$_{\rm meas}$($^{26}$Al) &
$(\frac{^{26}{\rm Al}}{^{27}{\rm Al}})_{\rm exp}$ [$\times10^{-11}$] & $(\frac{^{26}{\rm Al}}{^{27}{\rm Al}})_{\rm ref}$\footnotemark[1] [$\times10^{-11}$] & $(\frac{^{26}{\rm Al}}{^{27}{\rm Al}})_{\rm abs}$\footnotemark[2] [$\times10^{-11}$] & N$_{\rm abs}$($^{26}$Al)\footnotemark[2]$^,$\footnotemark[3] [$\times10^6$]\\
\hline
V1          & 3700 & 77 & 26668 & $1.50\pm0.01$  & $1.62\pm0.03$ & & \\
M11         & 3700 & 86 & 17394 & $0.875\pm0.007$  & $ 1.00\pm0.02$ & & \\
Al$_2$O$_3$ & 3700 & 101 & 17 & $0.0007\pm0.0002$  & & & \\
304-BLK     & 3700 & 20 & 11 & $0.0008\pm0.0002$  & & & \\
304-S1      & 3700 & 80 & 16769 & $0.906\pm0.008$  & & \multirow{2}{*}{$1.01\pm0.02$\footnotemark[4]}	 &	 \multirow{2}{*}{$225\pm4$}\\
304-S2      & 3700 & 61 & 12602 & $0.893\pm0.007$  &  \\
\hline
M11	        & 6400 & 79 & 26077 & $0.825\pm0.007$ & $1.00\pm0.02$ & & \\
304-S3     & 10400 & 21.4 & 11478 & $0.825\pm0.005$ & & $1.00\pm0.02$ & $223\pm5$ \\
304-R1     & 10400 & 22.6 & 327   & $0.0223\pm0.0002$ & & $0.0270\pm0.0006$	& $3.02\pm0.07$\\
\end{tabular}
\end{ruledtabular}
\footnotetext[1]{from Ref. \cite{Wallner}}
\footnotetext[2]{The uncertainty includes the statistical error and the accuracy of the reference samples.}
\footnotetext[3]{The uncertainty of the $^{27}$Al carrier weight was added quadratically.}
\footnotetext[4]{Average value from the two samples 304-S1 and 304-S2.}
\end{minipage}
\end{table*}

\subsection{AMS measurement of $^{26}$Al/$^{27}$Al isotopic ratios}

The details of the AMS system are published elsewhere \cite{CIRCE} while the chemical preparation of the sample cathodes is described in Appendix \ref{chemistry}. The $^{26}$Al/$^{27}$Al isotopic ratio was obtained measuring the $^{27}$Al current in a Faraday cup and the $^{26}$Al ions in a Silicon detector at the final focal plane. A high voltage applied to the chamber of the analyzing magnet allowed a fast switching between $^{26}$Al and $^{27}$Al measurements.

The obtained isotopic ratio ($^{26}$Al/$^{27}$Al)$_{\rm exp}$, however, depends on the experimental conditions of the AMS system and need to be normalized to a reference sample. The isotopic ratios ($^{26}$Al/$^{27}$Al)$_{\rm ref}$ of the reference samples V1 and M11 are well known from other experiments with a precision of 2\% \cite{Wallner}. The comparison of ($^{26}$Al/$^{27}$Al)$_{\rm exp}$ and ($^{26}$Al/$^{27}$Al)$_{\rm ref}$ for the reference samples lead to a correction factor which has to be applied to the experimental isotopic ratios of the $^{25}$Mg samples in order to evaluate their absolute isotopic ratios ($^{26}$Al/$^{27}$Al)$_{\rm abs}$. Finally, the known amount of $^{27}$Al added during the sample preparation (see Table \ref{AMS-samples}) allows together with the absolute isotopic ratios for calculating the total number of $^{26}$Al nuclei in the sample and, in turn, the resonance strength for the $^{25}$Mg(p,$\gamma$)$^{26}$Al resonance (see below).

The best overall efficiency, i.e. the number of $^{26}$Al detected with respect to the $^{26}$Al pressed into the cathode, was about $2\times10^{-4}$. Table \ref{AMS_results} shows the results of the AMS measurements for the various samples performed at the CIRCE laboratory. The experimental ratios of 304-S1, 304-S2, and 304-S3 are compatible within the statistical errors proving the reproducibility of the AMS measurements. In addition, we report the isotopic ratios of blank cathodes filled with a) standard aluminum oxide (sample Al$_2$O$_3$) and b) the material resulting from the chemical procedure applied to the non-irradiated $^{25}$Mg target C (sample 304-BLK). The results of these two blank samples are in perfect agreement confirming that no additional $^{26}$Al background is present in the $^{25}$Mg targets. The $^{26}$Al background level was equal to a isotopic ratio of $8\times10^{-15}$ and, thus, negligible with respect to the counting rate of the irradiated samples.

\subsection{Results and discussions}

The ground state resonance strength $\omega\gamma_{gs}= f_0\cdot\omega\gamma$ of the ${\rm E}=304$~keV resonance in $^{25}$Mg(p,$\gamma$)$^{26}$Al can be evaluated from eq. (\ref{omegagamma}). The factor $f_{0}$ is the probability for forming the ground state of $^{26}$Al which in the present experiment was determined with the HPGe detector (section \ref{HPGe_strength}).
However, the stoichiometry of the targets was not determined independently, therefore an independent evaluation
of $\omega\gamma_{gs}$ of the ${\rm E}=304$ keV resonance could not be done. Nevertheless, the comparison between the two independent, relative results, i.e. AMS and BGO $\gamma$-ray spectroscopy, is testing the precision of both methods. The ratio between the two methods can be determined with high precision by the relation:
\begin{equation}\label{ratio}
\frac{(\omega\gamma_{gs})_{\rm AMS}}{(\omega\gamma_{gs})_{\rm BGO}}= \frac{\rm Y_{^{26}{Al}}}{f_0\cdot{\rm Y_\gamma}} = \frac{\rm N_{abs}(^{26}{Al})}{f_0}\cdot \frac{\epsilon_{\rm BGO}}{\rm N_\gamma}
\end{equation}
where Y$_{^{26}{\rm Al}}$ and Y$_\gamma$ are $^{26}$Al and $\gamma$-ray yield per incident projectile. The parameter N$_\gamma/\epsilon_{\rm BGO}$ can be evaluated with the {\scriptsize GEANT4} simulation from the related BGO $\gamma$-ray spectra, analyzed with the same procedure described previously. The different dead time for both yield values need to be taken into account.
The AMS measurement has an uncertainty of 2\% arising from the normalization to the reference samples \cite{Wallner}. Additional systematic errors take into account the uncertainties of the carrier weight (0.2\%) and the efficiency of the chemical treatment of the samples (1\%). This leads to an overall systematic uncertainty of 2.5\% for the AMS measurement. The uncertainties of the corresponding $\gamma$-ray measurement are given by the absolute efficiency determination for the BGO detector (3\%), the decay scheme (3\%), and the ground state probability $f_{0}$ (1\%): a total systematic error of 4\%. The uncertainty of the effective stopping power and the charge integration represent common uncertainties to both methods and, therefore, were not taken into account in the error estimate for the $\omega\gamma$ ratio.

\begin{table}
\caption{Comparison between AMS result and BGO $\gamma$-ray measurement.}\label{AMS_comparison}
\begin{ruledtabular}
\begin{tabular}{c c c c}
  & AMS & BGO prompt $\gamma$-ray & \multirow{2}{*}{$\frac{(\omega\gamma_{gs})_{\rm AMS}}{(\omega\gamma_{gs})_{\rm BGO}}$} \\
Target	& N$_{\rm ^{26}Al}$ [$\times10^{6}$]\footnotemark[1] & N$_\gamma \cdot f_{0}/\epsilon_{\rm BGO}$ [$\times10^{6}$]\footnotemark[2]  &  \\
\hline
304-S	& $224\pm7$	& $218\pm9$	& $1.03\pm0.03$\footnotemark[3] \\
304-R	& $3.02\pm0.12$	& $3.02\pm0.13$	& $1.00\pm0.04$\footnotemark[3] \\
average & & & $1.02\pm0.05$\footnotemark[4] \\
\end{tabular}
\end{ruledtabular}
\footnotetext[1]{includes systematic uncertainty of 2.5\% (see text).}
\footnotetext[2]{includes systematic uncertainty of 4\% (see text).}
\footnotetext[3]{statistical uncertainties only}
\footnotetext[4]{systematic uncertainty not common to both methods were added quadratically}
\end{table}

The results are shown in Table \ref{AMS_comparison}. The ratio of $(\omega\gamma_{gs})_{\rm AMS}/(\omega\gamma_{gs})_{\rm BGO}=1.02\pm0.05$ is clearly consistent with unity which demonstrates that there is no major systematic uncertainties in the efficiency determination of the present $\gamma$-ray measurements. Furthermore, a high quality AMS measurement is possible and no systematic difference exists between AMS result and $\gamma$-ray data.

\section{Summary and consequences}\label{discussion}

In the present work we have measured properties (strengths $\omega\gamma$ and branching ratios) of the  ${\rm E}=214$, 304, and 326 keV resonances in the reactions $^{24}$Mg(p,$\gamma$)$^{25}$Al, $^{25}$Mg(p,$\gamma$)$^{26}$Al, and $^{26}$Mg(p,$\gamma$)$^{27}$Al, respectively. These new results together with selected previous work (see section \ref{gamma-discussion}) are used to calculate updated recommended values for the resonance strengths (Table \ref{omegagamma_comparison}). We underline that the new results were obtained from measurements with partly independent approaches, yielding a remarkable agreement among them. This is a strong evidence for the internal consistency of the present approach and demonstrates the achieved accuracy. Moreover, particular attention was paid to the critical issue of target stoichiometry and its variation under beam bombardment: a severe problem for Mg targets.

The reduction of the uncertainties of these Mg-Al cycle reactions is, by itself, an important improvement in the analysis of the nucleosynthesis scenarios relevant for the production and destruction of Mg and Al isotopes. A detailed discussion of the astrophysical implications is beyond the scope of the present work and will be published elsewhere. Nevertheless, the particular impact on each reaction rate and our recommendations are discussed in the following.

\begin{table}
\caption{Summary of the new recommended resonance strength values obtained as weighted average from present and previous work as discussed in section \ref{gamma-discussion}.}\label{omegagamma_comparison}
\begin{ruledtabular}
\begin{tabular}{c c}
 reaction and resonance                           & $\omega\gamma$ [meV] \\
\hline
 $^{24}$Mg(p,$\gamma$)$^{25}$Al ${\rm E}=214$ keV & $11.2\pm0.9$ \\
 $^{25}$Mg(p,$\gamma$)$^{26}$Al ${\rm E}=304$ keV & $30.8\pm1.3$ \\
 $^{26}$Mg(p,$\gamma$)$^{27}$Al ${\rm E}=326$ keV & $271\pm10$ \\
\end{tabular}
\end{ruledtabular}
\end{table}

The $^{25}$Mg(p,$\gamma$)$^{26}$Al resonance strength given by previous compilations, NACRE \cite{angulo}, or recent evalunations \cite{ReacRate1} is basically confirmed by the present experiment. Therefore, the nuclear reaction rate recommended in these compilations needs only a minor adjustment and is not recalculated here. We suggest to use the existing compilations until the results for low-energy $^{25}$Mg(p,$\gamma$)$^{26}$Al resonances, i.e. the resonances at ${\rm E}=93$ and 190~keV, will be available. However, the uncertainty of the ${\rm E}=304$ keV resonance could be reduced to 4\%. This is important since this resonance will serve as a normalization standard for the further measurements at low-energies and, thus, these results will benefit strongly from the present work. Furthermore, the primary branching ratios have been measured with high precision yielding a ground state feeding probability, $f_0=87.8\pm1.0\%$.

An additional AMS measurement based on an irradiation performed simultaneously to a $\gamma$-ray detection showed no systematic difference between both detection techniques. The different result of Arazi et al. \cite{arazi} was not observed in the present experiment and most likely caused by an unidentified oxygen contamination in the sample during the irradiation. In general, the normalization of the AMS measurement, e.g. the irradiation, is a challenging experimental task. In the present experiment the standard approach, i.e. resonance scans of the thick-target yield, appeared to be sufficient. Certainly, at lower energies - in particular for the ${\rm E}=93$ and 190~keV resonances - this approach cannot be applied due to the production of additional $^{26}$Al nuclei yielding a sizeable fraction of the total $^{26}$Al amount. However, such a normalization is of utmost importance for a reliable AMS measurement. Alternatively, the $^{26}$Al yield can be monitored with the ${\rm E}=214$ keV resonance of $^{24}$Mg(p,$\gamma$)$^{25}$Al. This resonance is located in an energy window where no additional $^{26}$Al production is expected.

The case of $^{24}$Mg(p,$\gamma$)$^{25}$Al is rather complex. The present recommended $\omega\gamma$ of the lowest resonance, ${\rm E}=214$, in this reaction is lower by more than 10\% compared to the latest published measurement by Powell et al. \cite{PO99}. However, as mentioned in the introduction a strong direct capture component dominates the resonance contribution and a reanalysis of the reaction at astrophysical energies by using an R-matrix formalism may prove worthwhile. Moreover, as demonstrated in \cite{PO99} the narrow resonance approximation \cite{Rolfs88} cannot be applied to evaluate the reaction rate on the low-energy tail of this resonance. Hence, the $\gamma$-width $\Gamma_\gamma$ need to be known to a high precision which is presently not achieved. A detailed study of the influence of all parameters involved is far beyond the scope of the present study and therefore postponed to a future work. As a consequence we do not give an updated nuclear reaction rate for the $^{24}$Mg(p,$\gamma$)$^{25}$Al reaction at the present stage.

The reaction $^{26}$Mg(p,$\gamma$)$^{27}$Al proceeds very fast at all temperatures compared to the other Mg-Al cycle reactions and, therefore, its astrophysical implications are negligible. Nevertheless, an apparent discrepancy in the literature (see for example \cite{angulo}) has been solved and the strength of the ${\rm E}=326$ keV $^{26}$Mg(p,$\gamma$)$^{27}$Al resonance was measured with a high precision (Table \ref{omegagamma_comparison}).

\begin{acknowledgments}

The authors are grateful to H.~Baumeister (Institut f\"ur Kernhysik, Westf\"alische Wilhelms Universit\"at M\"unster, Germany) and M. Lorrigiola (Laboratori Nazionali di Legnaro, Padova, Italy) for the excellent preparation of natural and enriched Mg targets. The present work has been supported by INFN and in part by the EU (ILIAS-TA RII3-CT-2004-506222), OTKA (T49245 and K68801), and DFG (Ro~429/41).
\end{acknowledgments}

\appendix
\section{Chemical extraction of $^{26}$A\MakeLowercase{l} from the $^{25}$M\MakeLowercase{g} bulk}\label{chemistry}

Aluminum can be extracted from the sputter ion source either as negative Al$^-$ ions or oxide molecules AlO$^-$. In spite of the higher extraction efficiency (about a factor 20 \cite{arazi}) for AlO$^-$ in the present measurement the Aluminum was injected as Al$^-$, since Mg does not form negative ions, while MgO does. Hence, isobar interferences were avoided in the AMS measurement. As a consequence all the target material removed from the target backing could in principle be used for preparing the sputter cathode. However, the amount of material needed for a single cathode is very small and, thus, a reduction of the Mg bulk material, i.e. a purification process, was necessary. As a first step a stoichiometric amount of $^{27}$Al - serving as a carrier - had to be added to the target material before the chemical extraction. The standard extraction procedure is based on the precipitation of the Al as hydroxide using ammonia followed by ignition of the precipitate to educe aluminum oxide:
							$$ {\rm AlCl_{3} + NH_{4}OH \rightarrow Al(OH)_{3} + 3NH_{4}Cl} $$
This procedure is limited because: i) The aluminum hydroxide precipitates as a gel which strongly retains the mother solution and purification as well as manipulation become difficult. ii) The $^{26}$Al yield scales with the reagent concentrations. In order to extract reasonable yields a high $^{26}$Al concentration is needed implying very small total volumes. Small volumes are difficult to handle. iii) After dehydration the final product (Al$_2$O$_3$) is a fine powder adhering to the walls and, thus, increase the risk of losses during the cathodes preparation.

Therefore, a new improved procedure, based on liquid-liquid extraction of an organic aluminum complex, has been developed and optimized in the LNGS chemistry laboratory. The Al reacts with three 8-hydroxyquinoline molecules to form a coordination compound insoluble in water but highly soluble in chlorinated organic solvents so that it can be extracted and separated. The reaction product can be converted to Aluminum oxide by heating to high temperature. The main advantages of the liquid-liquid extraction are:
\begin{itemize}
\item The handling is easier because of larger liquid volumes.
\item The yields - measured with ICPMS (inductively-coupled-plasma mass spectrometry) technique - are fully reproducible and found to be $>99$\% of the initial value.
\item The remaining magnesium concentration is lower than 1\% of the original value with a proper control of the pH value.
\item The final product, Al$_2$O$_3$, is obtained in a well crystalized form, not adhering to the vial walls.
\end{itemize}

After the chemical extraction, the mixture of $^{26}$Al and $^{27}$Al material was pressed together with a comparable amount of Cu powder in Cu cathodes.

\bibliographystyle{apsrev4-1}

\end{document}